\documentclass{ccjnl}
\usepackage{multirow}

\title{Far-Field vs. Near-Field Propagation Channels: Key Differences and Impact on 6G XL-MIMO Performance Evaluation}
\author{Zihang Ding\inst{1}, Jianhua Zhang\inst{1,*}, Changsheng You\inst{2}, Pan Tang\inst{1}, Hongbo Xing\inst{1}, Zhiqiang Yuan\inst{3}, Jie Meng\inst{1}, Guangyi Liu\inst{4}\corinfo{jhzhang@bupt.edu.cn}}
\receiveddate{XX}
\reviseddate{XX}
\Editor{XX}

\address[1]{State Key Lab of Networking and Switching Technology, Beijing University of Posts and Telecommunications, Beijing 100876, China}
\address[2]{Department of Electrical and Electronic Engineering, Southern University of Science and Technology, Shenzhen 518055, China}
\address[3]{National Mobile Communications Research Laboratory, School of Information Science and Engineering, Southeast University, Nanjing 210096, China}
\address[4]{China Mobile Research Institute, Beijing 100053, China}

\begin{document}
\maketitle
\begin{abstract}
Extremely large-scale multiple-input multiple-output (XL-MIMO) is regarded as a promising technology for next-generation communication systems. However, this will expand the near-field (NF) range, rendering more users more likely to be located in the NF region. In this paper, we aim to answer two questions: What are the new characteristics of the NF channel? Is it necessary to develop new transciver techniques to maintain system performance within the NF region? To this end, we first review current NF channel models and analyze the differences between the existing 3GPP TR 38.901 channel model and the NF channel model, including the spherical wavefront and spatially non-stationarity. Then, we provide examples on how these differences affect the XL-MIMO system performance in terms of beamforming gain and achievable rate. Simulation results demonstrate that, when using far-field (FF) technique under the NF channel, the maximum normalized beam gain loss is less than 3 dB for most users in the NF region defined by Rayleigh distance. Moreover, the achievable rate loss of beam training is less than 3\% compared to that realized by NF technique. Finally, we demonstrate the necessity of employing NF transceiver techniques based on simulation results.
\keywords{Near-field communication; channel model; near-field effect; spatial non-stationary}
\end{abstract}

\section{Introduction}
The sixth generation (6G) mobile communication systems are expected to facilitate immersive communication, provide ubiquitous connectivity, and lay a solid foundation for intelligent interconnection of all things in the future \cite{ITU-RM.2160,10121509}. To meet the diverse application demands of future 6G services, the use of high-frequency bands with large available bandwidths is required \cite{liu2023beginning,zhang2023channel,lu2024tutorial}. However, high-frequency band communications encounters challenges such as severe path loss and limited coverage range. One potential solution is to combine high-frequency communication with the beamforming technology of extremely large-scale multiple-input multiple-output (XL-MIMO) systems \cite{cui2022near,liu2023vision}. On the one hand, high frequencies enable the integration of more antenna elements within a limited space in XL-MIMO systems; on the other hand, the beamforming capability of these extremely large-scale antenna arrays can concentrate the energy of high-frequency signals into directional narrow beams for data transmission. This not only compensates for the path loss incurred by high frequencies and enhances the transmission distance of signals, but also enables spatial multiplexing of spectrum resources, thereby improving spectrum utilization efficiency \cite{6829967,7939954,rappaport2019wireless}. However, the increase in antenna aperture expands the near-field (NF) range, rendering more users more likely to be located in the NF region, and the impact of the NF can no longer be neglected \cite{cui2024near,liu2023near,tang2024xl}.

For wireless communication systems, the channel between the transmitter and receiver determines the communication performance upper-bound (e.g., channel capacity) \cite{molisch2012wireless,9312675}. Moreover, channel modeling has wide applications in the design of communication systems, such as verifying of new technology performance, deployment and optimization of communication networks, and testing of terminal performance. Therefore, for XL-MIMO systems, a comprehensive understanding and accurate modeling of the wireless channel are of paramount importance that support their research and technological advancement \cite{zhang2024new}. Compared to traditional small array systems, XL-MIMO systems are typically equipped with extremely large-scale antenna arrays and operate at high-frequency with wide bandwidths. Hence, it can no longer approximate the spherical wavefront propagation of electromagnetic waves under the traditional far-field (FF) assumption of planar wavefront propagation in XL-MIMO systems \cite{liu2023near,zhang20236g}. The nonlinear change in signal phase and differences in propagation distances between antenna elements must be considered, rendering XL-MIMO channels to exhibit NF propagation and spatial non-stationarity (SnS) characteristics \cite{cui2022near,liu2023near,lu2024tutorialnearfieldxlmimocommunications,yuan2022spatial,10683499}. Leveraging novel signal processing techniques based on these new characteristics, XL-MIMO is expected to significantly enhance system performance. However, on the other hand, these new characteristics also introduce new challenges for channel modeling, performance analysis, and system design in the NF region \cite{lu2024tutorial,tang2024xl}. 
For example, for near-field beam training, it is necessary to consider how to reduce the additional training overhead introduced by NF spherical wavefronts \cite{11,22,33,44,55}. Based on the aforementioned discussions, two natural questions arise: What are the fundamental differences between NF and FF propagation channels, and how to model these new channel characteristics? Furthermore, how these characteristics affect system communication performance in practice?
		
In this paper, we provide an overview of the research progress in NF channel modeling and summarize the key differences between NF and FF channels, as well as explore their impact on the performance of future 6G XL-MIMO systems. Numerical results reveal that within the NF region defined by the Rayleigh distance \cite{selvan2017fraunhofer}, the maximum normalized beam gain loss caused by the FF technique in NF channel is less than 3 dB for the majority of users. Notably, severe performance losses occur when the distance is less than approximately 0.1 Rayleigh distance. Additionally, compared to employing NF technique in NF channel, the impact of using FF technique on the achievable rate of beam training is less than 0.3 bps/Hz under different transmit signal-to-noise ratios (SNRs). Therefore, the decision to utilize NF techniques should be targeted and contingent upon specific objectives.

\section{Near-field Propagation Channel Characteristics}
In this section, we first introduce the definition of the NF and then focus on the recent advancements in NF channel modeling research. Finally, we provide a comparative summary of two representative channel models: the traditional FF 3GPP TR 38.901 channel model framework and a NF channel model framework based on the capturing and observation of multipath propagation mechanisms \cite{yuan2022spatial,3gpp.38.901}.

\subsection{The Definition of Near-Field}
\begin{figure}[]
	\centering	\includegraphics[width=\linewidth]{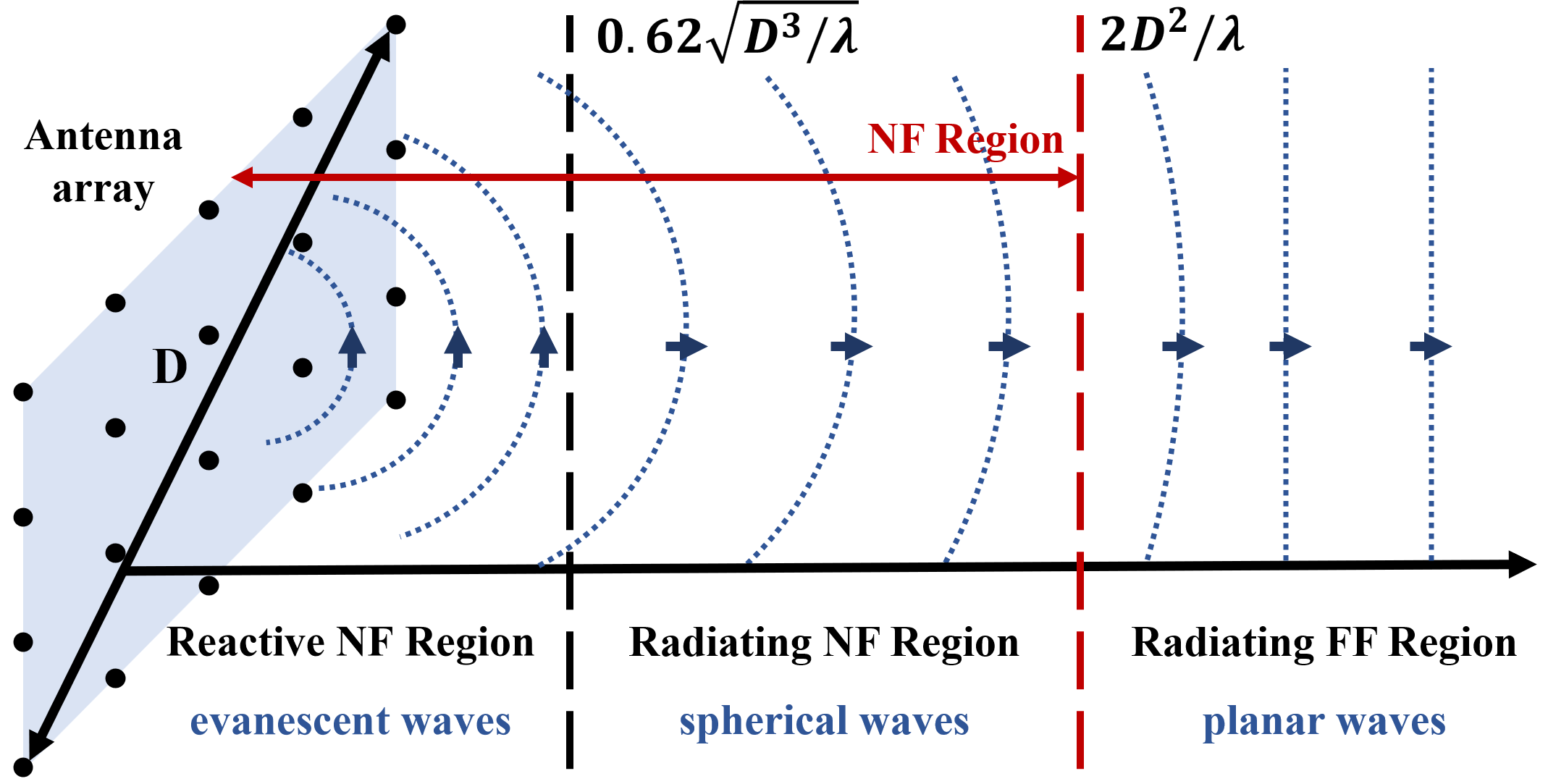}
	\caption{The definition of field regions for an antenna array.}
	\label{fig:2.1}
\end{figure}

According to electromagnetic and antenna theory, the electromagnetic field radiated by an antenna array can be divided into three distinct regions based on the characteristics of radiation propagation \cite{selvan2017fraunhofer}, as depicted in Figure \ref{fig:2.1}.

\subsubsection{Reactive NF Region}
In this region, the electromagnetic field varies with time, and the electric and magnetic field components are out of phase. This misalignment prevents the electromagnetic energy from radiating outward; instead, it continuously oscillates within the region. As a result, non-propagating fields known as evanescent waves dominate the reactive NF region. These waves exhibit a rapid exponential decay in power as the distance from the antenna increases, thereby confining their effectiveness strictly to the reactive NF region.

\subsubsection{Radiating NF Region (Fresnel Region)}
In this region, the electric and magnetic field components are perpendicular and in phase, enabling the emergence of propagating waves where the electromagnetic energy can radiate outward. However, the angular distribution of the field (beam pattern) is dependent on the distance from the antenna to the user, which results in spherical wavefronts.

\subsubsection{Radiating FF Region (Fraunhofer Region)}
In this region, the amplitude of the field diminishes inversely proportional to the distance from the antenna to the user, and the beam pattern becomes essentially independent of the distance difference, resulting in planar wavefronts. Consequently, the signal paths between different elements on the array and the user can be considered as parallel to each other.

In this paper, we distinguish between the NF and FF for XL-MIMO using classic boundaries, while the current boundaries being comprehensively outlined in \cite{sun2023differentiate}. The boundaries of different regions in this paper are indicated in Figure \ref{fig:2.1}, where $D$, and $\lambda$ represent the aperture of the antenna array and the wavelength of the carrier wave, respectively. 

Due to the significant phase difference between the electric field and magnetic field in the reactive NF region, the imaginary part of the electromagnetic field's Poynting vector dominates. Therefore, energy exchanges between the electric field and the magnetic field back and forth, and the energy of the electromagnetic field oscillates internally in the reactive NF region. In addition, the intensity of the electromagnetic field decays very rapidly with distance, so the reactive NF region usually cannot radiate signals outward, and the radiating NF region is typically more relevant to communication. Hence, we mainly consider the radiating NF region and contrast it with the radiating FF region. Unless specified otherwise, radiating NF is referred to as NF for simplicity.

\subsection{Near-Field Channel Modeling}
As the number of antenna elements and carrier frequencies are expected to increase significantly in 6G, the NF region of XL-MIMO will enlarge dramatically. For instance, a virtual uniform circular array (UCA) of 0.5 $\mathrm{m}$ radii was adopted in \cite{yuan2022spatial}, and the corresponding Rayleigh distance at 29 GHz is around 193 $\mathrm{m}$. Therefore, NF communication will become a fundamental component of future 6G, necessitating accurate channel modeling for NF propagation. Based on the different modeling approaches, existing NF channel modeling researches can be broadly categorized into statistical (or stochastic) channel models and deterministic channel models. It should be noted that the statistical channel models mentioned in this paper includes geometry-based stochastic models (GBSMs) and non-geometrical stochastic models. Table \ref{tab:channel_model} presents the summary of main NF channel models.

\begin{table*}[]
	\centering
	\caption{The summary of main NF channel models}
	\label{tab:channel_model}
	\begin{tabular}{p{0.35\columnwidth}<{\centering}|p{0.5\columnwidth}<{\centering}|p{0.15\columnwidth}<{\centering}|p{0.85\columnwidth}<{\centering}}    
		\hline
		\textbf{Modeling Method}  &   \textbf{Focused Channel Characteristic} &  \textbf{Reference} &  \textbf{Descriptions} \\
		\hline
		\multirow{8}{*}[-2.5ex]{\begin{tabular}{c} Statistical Modeling\end{tabular}}  &  	\multirow{3}*{Spherical Wavefront} & \cite{3gpp.38.901} & {\begin{tabular}{c} Recommended to propose \\ a  spherical wavefront  approximation for NF \end{tabular}} \\
		\cline{3-4}
    	~  &  	~ & \cite{ma2021non} & Empirically corroborated the approximation model in NF \\
		\cline{3-4}
		~  &  	~ & \cite{lopez2018novel} & Developed a second-order approximate NF channel model \\
    	\cline{2-4}
		~  &  	\multirow{3}*{SnS} &  \cite{zheng2022ultra} & Introduced a three-dimensional non-stationary GBSM  \\
		\cline{3-4}
				~  &  ~ &  \cite{liu2012cost} & Pioneered the VR concept \\
		\cline{3-4}
		~  &  ~ & \cite{gao2013massive} & Further implemented the VR concept into XL-MIMO \\
		\cline{2-4}
		~  &  \multirow{2}{*}[-1.5ex]{\begin{tabular}{c} Spherical Wavefront, SnS\end{tabular}} & \cite{li20193d} & {\begin{tabular}{c} Used various frequency bands and scenarios GBSM to model \end{tabular}} \\
		\cline{3-4}
		~  &  ~ & \cite{yuan2022spatial} & {\begin{tabular}{c} Proposed a NF channel framework based on the capturing \\ and observation of multipath propagation mechanisms \end{tabular}} \\
		\hline
		 \multirow{2}{*}[-1.5ex]{\begin{tabular}{c} Deterministic Modeling\end{tabular}}  & Path Loss, etc  & \cite{yao2017massive} & Implemented map-based models to conduct RT \\
		\cline{2-4}
		~  & Spherical Wavefront, SnS  & \cite{yuan2024efficient} & {\begin{tabular}{c} Proposed an RT-based NF channel model \\ with precision and efficiency \end{tabular}} \\
		\hline	
	\end{tabular}
\end{table*}

\subsubsection{Statistical Channel Models}
Mathematically-based statistical channel modeling typically targets channels of different environmental categories, facilitating the probabilistic reconstruction of channels across diverse environments through parameter distribution. In contrast to conventional small-array channels, XL-MIMO channels manifest NF propagation effects and SnS. These intrinsic attributes must be precisely represented in the channel modeling to guarantee the fidelity of channel reconstruction.

The extant researches have widely investigated the NF propagation traits of XL-MIMO in channel modeling. For example, the 3GPP TR 38.901 \cite{3gpp.38.901} standardized channel model recommended to propose a spherical wavefront approximation for NF channels, hypothesizing that signals emanate from a point source with a spherical wavefront towards the array. \cite{ma2021non} empirically corroborated the adequacy of this approximation model in portraying NF propagation characteristics. Additionally, \cite{lopez2018novel} developed a second-order approximate channel model based on the spherical wavefront (i.e., parabolic wavefront) to simulate XL-MIMO NF effects in both spatial and temporal domains, enhancing the modeling efficacy.

However, few researches have considered SnS in NF channel models \cite{zhang2018recent}. \cite{zheng2022ultra} introduced a three-dimensional non-stationary GBSM tailored for XL-MIMO systems. The COST 2100 model \cite{liu2012cost} pioneered the visibility region (VR) concept, delineating XL-MIMO SnS by confining multipath clusters activity within different limited areas for base station antenna elements. However, the reliance on planar wavefront assumption restricts the accuracy of the COST 2100 model in simulating SnS feature. Based on the VR concept, \cite{li20193d} modeled SnS and NF effects in various frequency bands and scenarios. However, these reliability of these models should be substantiated by real empirical data. \cite{gao2013massive} further introduced the VR concept into XL-MIMO, positing that only antennas within the VR can detect multipath clusters, with a variable gain modulated by the NF distance to account for power fluctuations within the VR. 

\begin{figure}[]
	\centering	\includegraphics[width=1.05\linewidth]{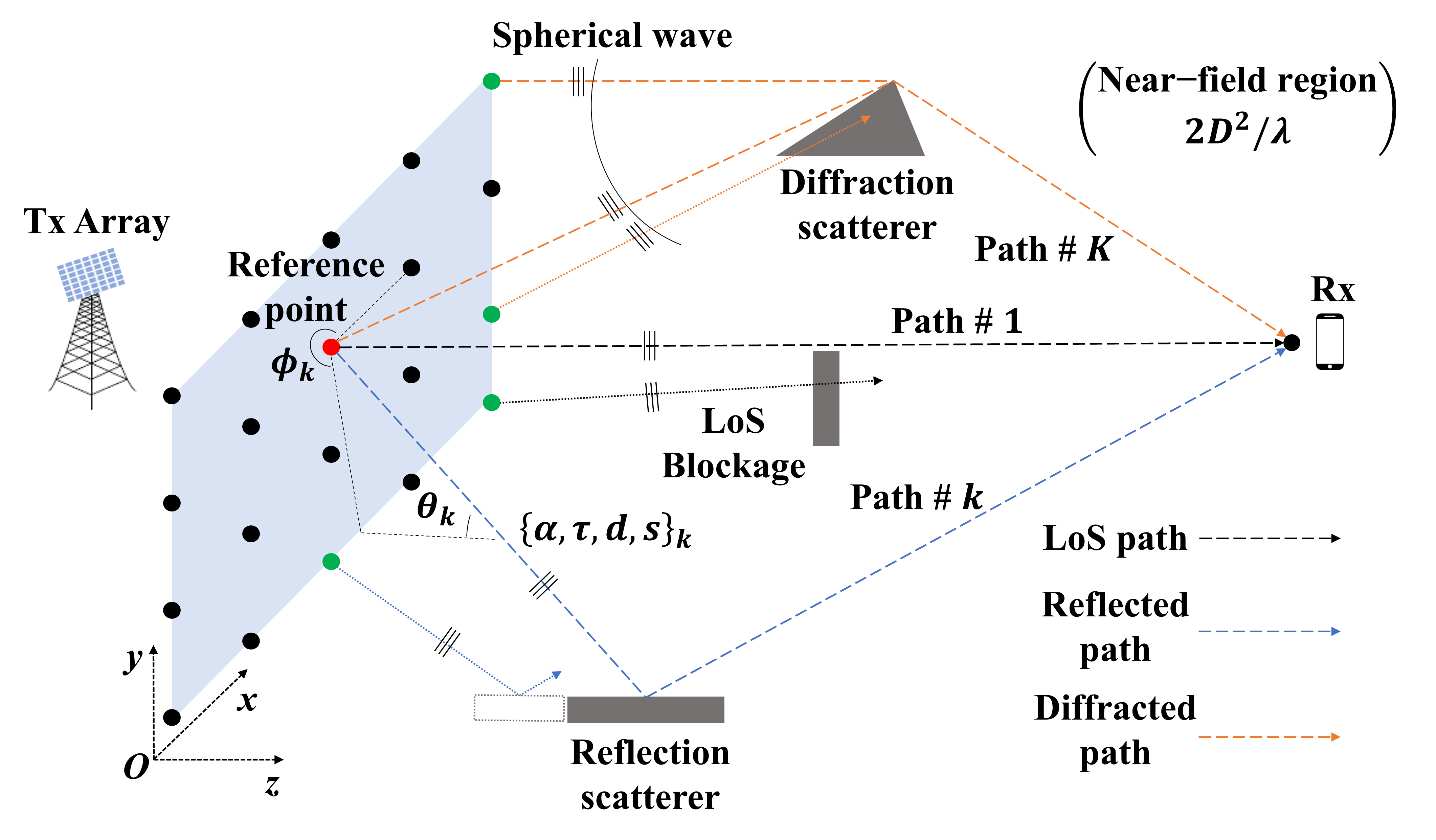}
	\caption{The spherical wavefront propagation with SnS in the NF region proposed in \cite{yuan2022spatial}.}
	\label{fig:2.2.1}
\end{figure}

To further characterize and elucidate SnS while reducing the complexity of XL-MIMO channel modeling, a NF framework based on the capturing and observation of multipath propagation mechanisms (i.e., LoS, reflection, and diffraction) was proposed in \cite{yuan2022spatial}, as shown in Figure \ref{fig:2.2.1}. There were $K$ SnS spherical wavefront multipath components between the transmit (Tx) array and user (Rx) in \cite{yuan2022spatial}. The XL-MIMO channel at frequency $f$ can be modeled as the superposition of the channel frequency responses (CFRs) of the $K$ paths on the array, which can be expressed as
\begin{equation}
\mathbf{H}^\mathrm{sns}(f)=\mathbf{S}\odot\mathbf{A}(f)\cdot\mathbf{H}(f),
\end{equation}
where $\mathbf{H}^\mathrm{sns}(f)\in \mathbb{C}^{M \times 1}$ comprises $M$ complex values, $\mathbf{S} \in \mathbb{C}^{M \times K}$, $\mathbf{A}(f) \in \mathbb{C}^{M \times K}$, $\mathbf{H}(f) \in \mathbb{C}^{K \times 1}$ denote the SnS characteristic matrix, the array manifold matrix and the CFRs at $f$ of the $K$ paths at the reference point. $f \in [f_{min},f_{max}]$ is the frequency range, and $\odot$ represents the elementwise product operation. The CFRs can be mathematically expressed as

\begin{equation}
\mathbf{H}(f)=[...,\alpha_ke^{-j2\pi f\tau_k},...]^\mathrm{T},
\end{equation}
where $\alpha_k$ and $\tau_k$ represent the complex amplitude and propagation delay of the $k$-th path, respectively. $(\cdot)^\mathrm{T}$ denotes the transpose operation. The manifold of $m$-th antenna element by the $k$-th path $a_{m,k}$ of the array manifold matrix can be represented by the propagation distance difference of the $m$-th element to the reference point:
\begin{equation}
a_{m,k}\left(f\right)=\frac{\left\|\boldsymbol{d}_k\right\|}{\left\|\boldsymbol{d}_{m,k}\right\|}e^{-j2\pi f\frac{\left\|\boldsymbol{d}_{m,k}\right\|-\left\|\boldsymbol{d}_k\right\|}{\mathbf{c}}},
\end{equation}
where $\| \cdot \|$ represents the Euclidean norm of the argument. $\boldsymbol{d}_k$ denotes the vector pointing from the reference point to the first scattering source of the $k$-th path propagation route. It is noteworthy that in the FF channel model, the propagation distance difference is approximated by the first-order expansion of a Taylor series. However, when considering the spherical wavefronts, this propagation difference is generally approximated by the second-order expansion of the Taylor series \cite{lu2024tutorial}.

The novel SnS characteristic matrix contains $K$ nonnegative real-valued vectors,
\begin{equation}
\mathbf{S}=[s_1,...,s_k,...,s_K],
\end{equation}
with
\begin{equation}
s_{k} = [s_{1,k},...,s_{m,k},...,s_{M,k}]^{\mathrm{T}},k\in [1,K],
\end{equation}
where $s_{m,k}$ is the SnS property of the $k$-th path on the $m$-th element.

In the traditional spatial stationary XL-MIMO channel model, the channel can be expressed as
\begin{equation}
	\mathbf{H}^\mathrm{ss}(f)=\mathbf{A}(f)\cdot\mathbf{H}(f),
\end{equation}
which can be regarded as a special case of the NF channel model proposed in \cite{yuan2022spatial}, with all $s_{m,k}$ equaling to 1. Consequently, compared to traditional statistical channel modeling, the proposed framework only introduced one additional parameter matrix for SnS. The spherical wavefront propagation with SnS characteristics in this NF model was based on measurements and ray tracing (RT) simulations, which achieving good real-time performance and high accuracy with low complexity.

\subsubsection{Deterministic Channel Models}
Deterministic channel modeling generates physically authentic channels sensitive to specific locations through environmental reconstruction and electromagnetic computations. It necessitates detailed information on the propagation environment of the scenario, the dielectric parameters of materials, and the precise geometric positions of Tx and Rx antennas. Focusing on the NF and SnS, this paper closely examines how existing deterministic channel modeling methods can accurately reconstruct XL-MIMO channels. RT is an extensively utilized method for deterministic channel modeling that employs simplified environmental reconstruction and electromagnetic field calculations to accurately simulate the wireless channels between any two spatial points of the Tx and the Rx \cite{fuschini2019study}.

The initial RT methods were based on simple point-to-point modeling to generate channels between the Tx and Rx \cite{karstensen2016comparison}. However, for multi-antenna systems, the simple RT modeling strategy is ineffective in generating a complete channel due to the large array aperture and the varying responses between array elements \cite{flordelis2019massive}. A straightforward strategy is to employ exhaustive modeling, where RT modeling is performed for each Tx/Rx antenna element pair. Nevertheless, for XL-MIMO channel modeling with hundreds to thousands of antenna elements, the computational complexity becomes unacceptable.

Therefore, the existing researches focus on how to reduce the complexity of RT channel modeling for multiple antennas. \cite{ng2007efficient} introduced a spatial partitioning strategy that excludes objects irrelevant to electromagnetic computations, thus accelerating the RT modeling process for multi-antenna configurations. Similarly, \cite{liu20203d}, \cite{tan2015full} had advanced techniques such as grid computing and parallel graphics processing unit (GPU) optimization to enhance the efficiency of RT modeling for multi-antenna systems. However, the aforementioned studies still use a comprehensive RT computation approach for multi-antenna channels, which might not be optimal for XL-MIMO channels. In \cite{yamada2009plane}, a novel RT method based on plane wavefront expansion to model MIMO channels was introduced, which generated the channel for one Tx-Rx antenna pair through RT and extended it to other pairs through plane wavefront propagation, but only suitable for small array systems. Following this, \cite{ng2007efficient}, \cite{arikawa2014simplified} utilized the plane wavefront expansion technique for MIMO channel generation and compared it with exhaustive RT modeling, revealing a substantial agreement in channel capacity. For XL-MIMO RT modeling, map-based models had been implemented to conduct RT by reducing the complexity of environmental geometry and electromagnetic description \cite{yao2017massive}. While such models may lack precision in channel prediction at specific locations, they are adept at simulating the SnS of XL-MIMO. 

\begin{figure}[]
	\centering	\includegraphics[width=\linewidth]{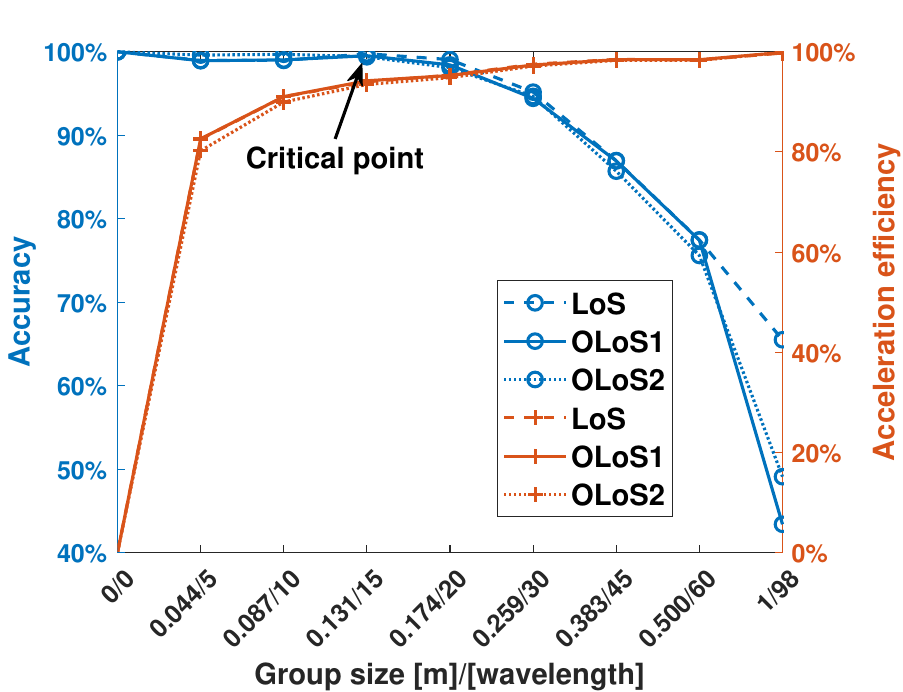}
	\caption{The analysis of the acceleration efficiency and accuracy in the coarse-refinement method.}
	\label{fig:2.2.2}
\end{figure}

Recently, \cite{yuan2024efficient} introduced an RT-based XL-MIMO channel model that captures NF
and SnS with precision and efficiency, which employs RT for a limited number of sparse antenna array elements and employs spherical/astigmatic-wavefront approximation and the uniform theory of diffraction for interpolation across other elements, thereby significantly simplifying the simulation process without compromising accuracy. As shown in Figure \ref{fig:2.2.2}, experimental validation across three distinct scenarios has confirmed that the accuracy of this model is comparable to brute-force methods, achieving approximately $99.5\%$ precision with an impressive $94.3\%$ improvement in efficiency.

\subsection{Key Differences between Far-Field and Near-Field Propagation Channels}
By reviewing the definitions of FF and NF regions and the advancements in existing channel model research, it becomes evident that the key differences between FF and NF channels can be summarized as the application of plane wavefront/spherical wavefront assumptions and spatially stationary/spatially non-stationary assumptions. Traditional FF 3GPP TR 38.901 channel models adopt the plane wavefront and spatially stationary assumptions, which fail to accurately describe NF characteristics. In contrast, a recently proposed NF channel model based on the capturing and observation of multipath propagation mechanisms can accurately depict the SnS and NF effects. Below is a summary and comparison of the key differences between FF and NF channels.
\begin{itemize}[leftmargin= *]
	\item \textbf{Plane wavefront/spherical wavefront assumption:} In FF channels, the plane wavefront assumption is used, where the amplitude of the electromagnetic waves incident on various antenna elements remains constant, and the phase is typically approximated using a first-order Taylor expansion. Consequently, the electromagnetic wave phase varies linearly with distance across the antenna elements. In NF channels, the maximum phase difference between antenna elements exceeds $\pi/8$, rendering the plane wavefront approximation inaccurate \cite{sun2023differentiate}. Therefore, in the NF region, the amplitude of the electromagnetic waves incident on the antenna elements no longer remains constant, and the phase must be calculated based on the spherical wavefront assumption using a second-order Taylor expansion. The phase exhibits a nonlinear relationship with the distance between the antenna elements.
	
	\item \textbf{Spatially stationary/spatially non-stationary assumption:} FF channels assume spatially stationary characteristics because the array aperture is small, allowing all antenna elements to observe the same environment with identical obstacles, thus exhibiting spatial stationarity. In NF channels, the large array aperture means that elements can no longer completely observe the environment of the same obstacles. Also, different antenna elements can observe different spatial channels, leading to SnS.
\end{itemize}

\section{The Impact of Near-Field Channel Modeling on XL-MIMO Performance Evaluation}
In this section, we simulate FF and NF channels based on the beijing university of posts and telecommunication channel model generator-international mobile telecommunications for 2030 (BUPTCMG-IMT2030) channel simulation platform \cite{zhang_buptcmg_imt2030}. Then, we analyze the impact of channel models on XL-MIMO system performance evaluation using uniform linear array (ULA) beamforming codebook.

\subsection{Channel Model Parameters Generation}
In this subsection, we employ the BUPTCMG-IMT2030 to generate channel parameters based on the traditional FF 3GPP 38.901 channel model and the NF model proposed in \cite{yuan2022spatial}. The accuracy of the channel simulation platform has been shown in \cite{gao20233gpp}, and the NF channel parameters generated thereby can capture the key differences between FF and NF channels: the spherical wavefront assumption and the SnS assumption.

The configuration parameters for this simulation are illustrated in Table \ref{tab:my_label}. We select the UMi scenario as defined by 3GPP TR 38.901 and obtain channel parameters at 15 GHz for performance evaluation. To ensure the equitable comparison of FF and NF channel differences, the distance between the base station (BS) and user terminal (UT) is incremented in steps of $1$ $\mathrm{m}$, transitioning from the NF region to the FF region. Meanwhile, for each set of simulations at identical distances, all UTs are randomly assigned to different directions to get average channel parameters.

\begin{table}[]
	\centering
	\caption{Simulation Configurations} 
	\label{tab:my_label}
	\begin{tabular}{p{0.4\columnwidth}<{\centering}p{0.5\columnwidth}<{\centering}}    
		\hline
		\textbf{Parameters}  &   \textbf{Values} \\
		\hline
		Scenario  & Urban Micro (UMi) \\
		\hline
		Propagation Condition  &  LoS/NLoS\\
		\hline
		Carrier frequency  &  15 GHz\\
		\hline		
		Bandwidth &	2 MHz\\	
		\hline
		Antenna Element Type  &	Omnidirectional / Vertical-polarized \\
		\hline
		Antenna Element	Gain  &  5 dB/5 dB \\
		\hline
		BS Configuration  & \makecell{Fixed Position\\ Equipped with 150-element ULA}\\
		\hline
		UT Configuration  &  \makecell{Over 4000 Randomly Drops\\ Equipped with a Single Antenna}\\
		\hline
		BS-UT Distance  &	5 m-80 m\\
		\hline	
	\end{tabular}
\end{table}

\begin{figure}[]
	\centering	\includegraphics[width=\linewidth]{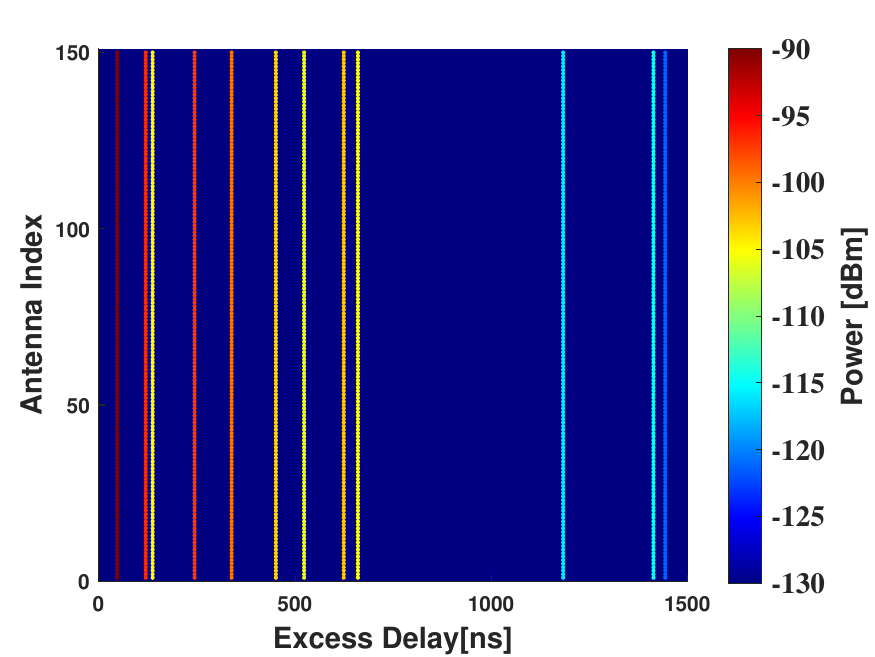}
	\caption{The heatmaps of PDPs in FF channel model.}
	\label{fig:3.1.1}
\end{figure}

\begin{figure}[]
	\centering	\includegraphics[width=\linewidth]{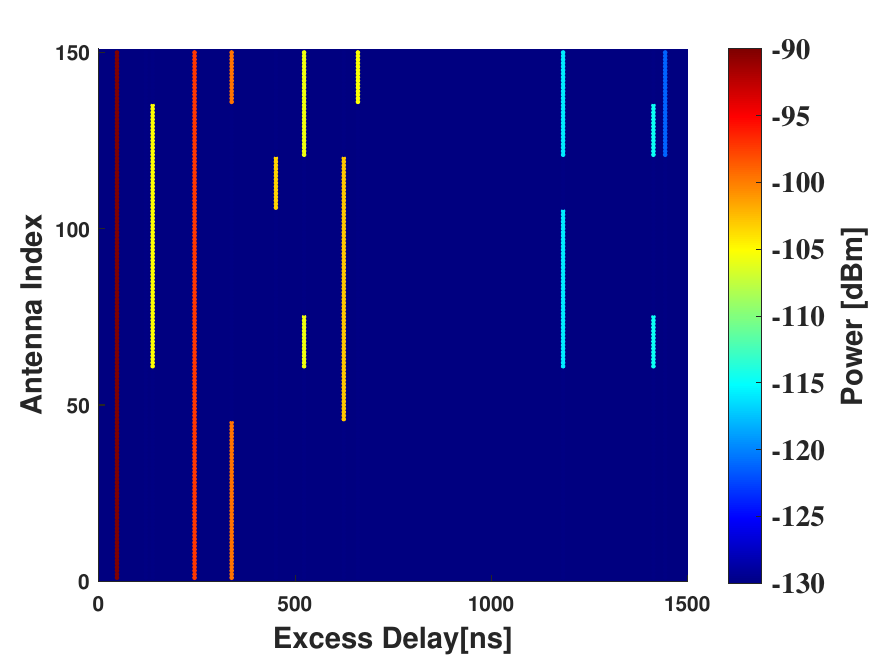}
	\caption{The heatmaps of PDPs in NF channel model.}
	\label{fig:3.1.2}
\end{figure}

Figure \ref{fig:3.1.1} and Figure \ref{fig:3.1.2} respectively illustrate the power delay profiles (PDPs) of the FF channel model 	$\mathbf{h}^\mathrm{FF}$ and the NF model $\mathbf{h}^\mathrm{NF}$ at a distance of $14$ $\mathrm{m}$, corresponding to a randomly selected direction from over 100 possible directions. It can be observed that in the FF channel model, each cluster is visible to all antenna elements, whereas in the NF channel model, clusters may appear or disappear across different antenna elements, indicating that the NF model captures the SnS characteristics.

\begin{figure}[]
	\centering	\includegraphics[width=\linewidth]{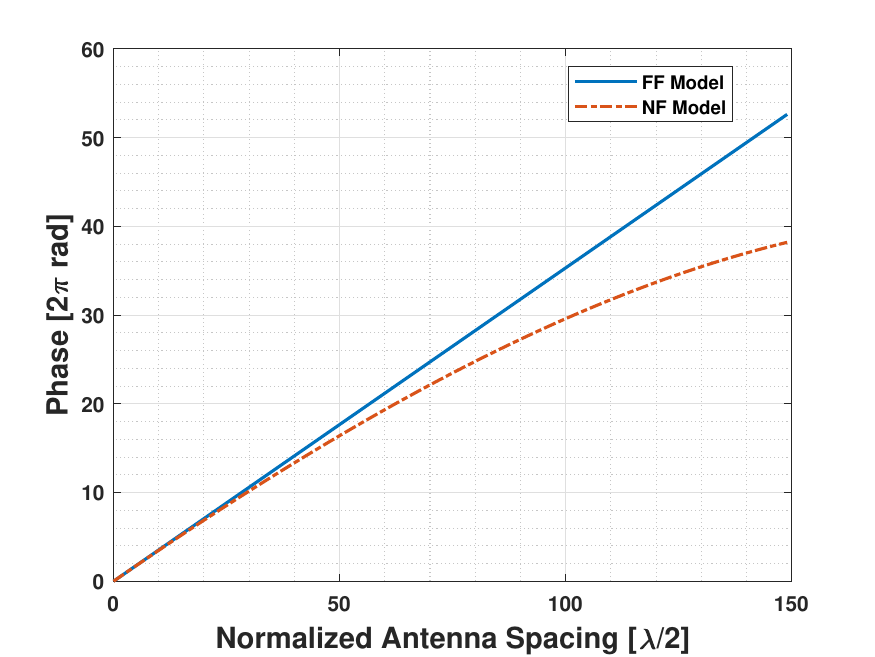}
	\caption{The phase variation of the LoS path versus the distance between antenna elements.}
	\label{fig:3.1.3}
\end{figure}

Figure \ref{fig:3.1.3} illustrates the relationship between the phase of the line-of-sight (LoS) path and the normalized antenna spacing of each BS antenna element to the reference point. In the FF channel model, the phase of the LoS path varies linearly with antenna spacing \cite{3gpp.38.901}. However, the NF channel model exhibits a nonlinear trend. This deviation is attributed to the consideration of the spherical wavefront in the NF channel model, where the phase is determined by the distance difference between the UT-BS antenna array elements and the UT-reference point. Since the distance changes nonlinearly, the phase of the LoS path also demonstrates a nonlinear variation. Furthermore, in the NF channel model, the same trend can be observed for the non-line-of-sight (NLoS) paths, indicating that the spherical wavefront applies to all paths within the NF channel.

\subsection{Performance Evaluation Comparison}
In this subsection, we employ the traditional FF ULA codebook and NF ULA codebook proposed in \cite{9693928} for beamforming and beam training, analyzing the impact of the differences between FF and NF channels on the system performance in terms of normalized beam gain and achievable rate.

\subsubsection{System Model}
We consider the downlink beam training in a narrowband communication system, where a base station (BS) equipped with $N$-antenna UCA serves a single-antenna user equipment (UE). The received signal at the UE can be expressed as
\begin{equation}
	y=\mathbf{h}^{\mathrm{H}}\mathbf{w}s+n_\sigma, 
\end{equation}
where $\mathbf{h}\in\mathbb{C}^{N\times1}$, $\mathbf{w}\in\mathbb{C}^{N\times 1}$, $s\in\mathbb{C}$, $n_\sigma\sim\mathcal{CN}(0,\sigma^2)$ denote the channel frequency response, codeword, transmitted signal, and additive white Gaussian noise, respectively.

We focus on the LoS channel between BS and UE, which is reasonable as we mainly rely on the dominant LoS path for data transmission in the millimeter-wave and even terahertz channels. As such, the LoS channel can be modeled as \cite{wu2023enabling}
\begin{equation}
	\mathbf{h}_\mathrm{LoS}=\sqrt{N}\alpha\boldsymbol{a}(\theta,r),
	\label{equ:H_resp2}
\end{equation}
where $\alpha$ denotes the complex amplitude of the LoS path. $\theta\triangleq\sin\varphi$ denotes the spatial angle, where $\varphi\in[-\pi/2,\pi/2]$ denotes the physical angle of the LoS path. $r$ denotes the distance between the BS and UE. $\boldsymbol{a}(\theta,r)$ represents the channel steering vector, which can be expressed as
\begin{equation}
	\boldsymbol{a}(\theta,r)=\frac{1}{\sqrt{N}}\Big[e^{\frac{-j2\pi(r^{(0)}-r)}{\lambda}},\ldots,e^{\frac{-j2\pi(r^{(N-1)}-r)}{\lambda}}\Big]^{\mathrm{T}},
\end{equation}
where $r^{(n-1)}$ denotes the distance between the $n$-th antenna element and UE. $\lambda$ denotes the carrier wavelength. 

\subsubsection{FF and NF ULA Codebook}
For the traditional FF ULA codebook $\mathbf{W}^\mathrm{FF}=\{\mathbf{w}^\mathrm{FF}_1,\mathbf{w}^\mathrm{FF}_2,...,\mathbf{w}^\mathrm{FF}_N\} $, which is designed based on the planar wavefront assumption \cite{33}. The $n$-th codeword $\mathbf{w}^\mathrm{FF}_n$ towards the predefined direction $\theta_n \textcolor{blue}{\in[-1,1]} $ is given by
\begin{equation}
	\mathbf{w}^\mathrm{FF}_n=\frac{1}{\sqrt{N}}\Big[1,e^{-j\pi\theta_n},\ldots,e^{-j\pi(N-1)\theta_n}\Big]^{\mathrm{T}},
\end{equation}
where $\theta_n=\frac{2n-N-1}{N}, n\in \mathcal{N} \triangleq \{1,2,\ldots,N\}$ and $N$ is the total antenna elements number.

Due to the spherical wavefront consideration, the design of NF codebooks is jointly determined by both the user distance and direction. Hence, for NF ULA codebook  $\mathbf{W}^\mathrm{NF}=\{\mathbf{W}^\mathrm{NF}_1,\mathbf{W}^\mathrm{NF}_2,...,\mathbf{W}^\mathrm{NF}_N\} $ with $\mathbf{W}^\mathrm{NF}_n=\{\mathbf{w}^\mathrm{NF}_{n,0},\mathbf{w}^\mathrm{NF}_{n,1},...,\mathbf{w}^\mathrm{NF}_{n,S_n-1}\}, \forall n$ $\in$ $\mathcal{N} $, we consider the typical polar-domain NF codebook proposed in \cite{9693928}. Each codeword $\mathbf{w}^\mathrm{NF}_{n,s_n}, s_n$ $\in$ $\mathcal{S} _n$ $\overset {\Delta }{\operatorname* { \operatorname* { = } } }$ $\{ 0, 1, \ldots , S_n-1\}$ corresponds to a beam towards a predefined location $(\theta_n,r_{n,s_n})$, which is given by 
\begin{equation}
		\mathbf{w}^\mathrm{NF}_{n,s_n}=\boldsymbol{a}(\theta_n,r_{n,s_n}),
\end{equation} 
where $r_{n,s_n}=\frac{N^2d^2}{2s_n\beta^2\lambda}(1-{\theta_n}^2)$. $d=\lambda/2$ is the antenna spacing. $\beta$ is the column coherence function parameter. To obtain the distance sampling $r_{n,s_n}$, we set the column coherence of the near-field steering vector to be small enough, e.g., 0.5 in this paper, thereby setting $\beta$ to 1.6 according to reference \cite{9693928}.

\subsubsection{Normalized beam Gain}

\begin{figure}[]
	\centering	\includegraphics[width=\linewidth]{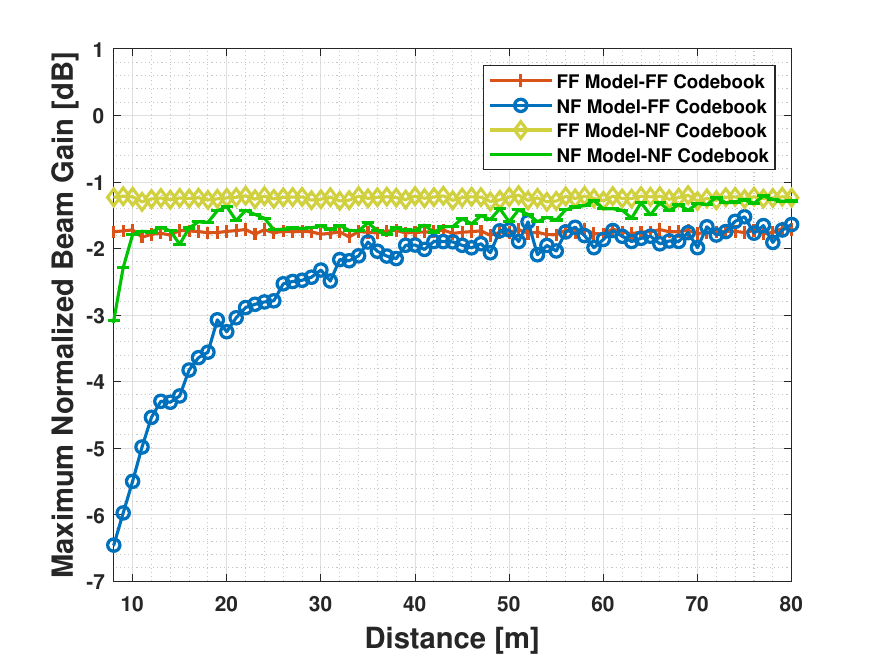}
	\caption{The maximum beamforming gain versus the distance between the BS and UT.}
	\label{fig:3.2.1}
\end{figure}

Figure \ref{fig:3.2.1} illustrates the maximum normalized beam gains $|{\boldsymbol{a}}^\mathrm{H}(\theta,r)\mathbf{w}|$ obtained at different distances between the BS and UT when different codebooks are used for different channel data. By comparing the FF and NF channel data beamformed by FF codebook (red and blue lines), as well as the NF channel data beamformed by the FF and NF codebooks (blue and green lines), it can be observed that, due to the introduction of spherical wavefronts and SnS characteristics in NF channels, there is a gain performance degradation of over 3 dB for FF codebook within the region where the distance between the BS and UT is less than about 22 $\mathrm{m}$ (approximately 0.1 Rayleigh distance), and the performance degradation becomes more severe as the distance decreases. On the contrary, using the NF codebook for NF channel can ensure effective coverage within the NF region (with gain loss less than 3 dB). Additionally, by comparing the FF and NF channel data beamformed by NF codebook (yellow and green lines), as well as the FF channel data beamformed by FF and NF codebooks (red and yellow lines), it can be observed that the NF codebook is still applicable for FF channel data. Furthermore, since the NF codebook additionally samples the distance domain while ensuring the same angular sampling, which incurs the additional distance sampling overhead, the maximum normalized beam gain ultimately achieved by NF codebook for FF channel data is higher than that achieved by the FF codebook.

Hence, it can be inferred that within most regions of the Rayleigh distance, using FF codebook for beamforming will not result in significant performance degradation. However, when the distance between the BS and UT is less than approximately 0.1 Rayleigh distance, NF codebook that accounts for spherical wavefronts and SnS characteristics is required to ensure effective coverage in the NF region.

\subsubsection{Achievable Rate}

\begin{figure}[]
	\centering	
	\includegraphics[width=\linewidth]{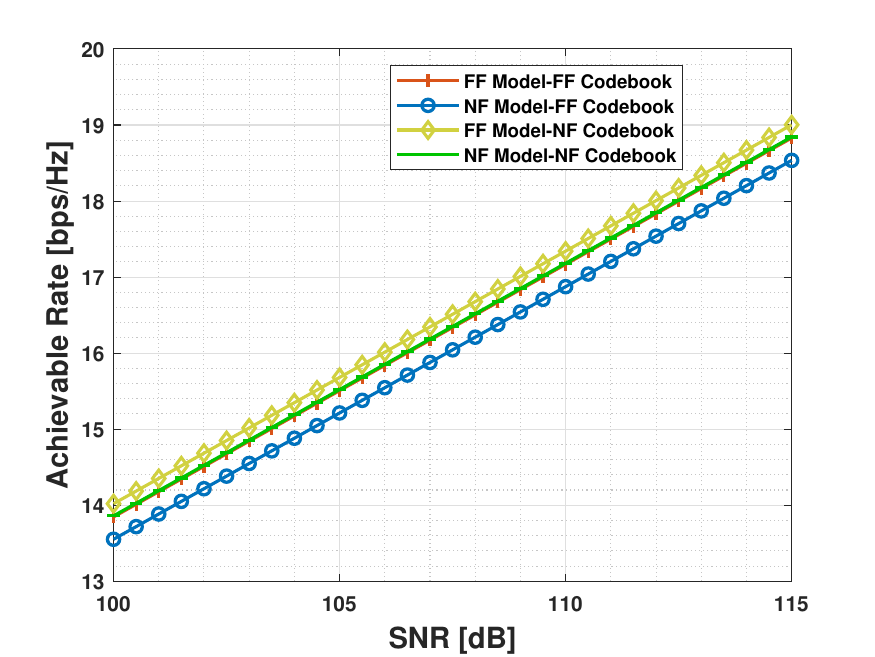}
	\caption{The achievable rate versus the transmit SNRs.}
	\label{fig:3.2.2}
\end{figure}

Figure \ref{fig:3.2.2} illustrates the achievable rates $R=\log_{2}\begin{pmatrix}1+\mathrm{SNR}_{\mathrm{tx}}\cdot|\mathbf{h}^{\mathrm{H}}\mathbf{w}|^{2}\end{pmatrix}$ versus the transmit SNRs through an exhaustive search method, with the transmit SNR ranging from +100 dB to +115 dB. It can be observed that the achievable rate of using FF technique for the NF channel data is consistently lower than that achieved by NF technique (blue and green lines), with a maximum difference of approximately 0.3 bps/Hz. Furthermore, the achievable rates of different channel datas treated with corresponding techniques are similar (red and green lines), and employing NF technique incurs additional training overhead to process FF channel data (yellow and red lines), resulting in the higher achievable rate. It is worth noting that the performance evaluation of training overhead is influenced by employing distinct beam training strategies with the same codebook. However, since the focus of this paper is on the impact of the channel on performance evaluation, a comparison of various NF beam training techniques is not provided. For research on different beam training strategies to reduce training overhead, please refer to e.g., \cite{11,33,10163797,10005200,Two}. Consequently, as low training overhead techniques are employed, it becomes imperative to employ NF techniques designed for NF channel steering vector to perform beam training.

\section{Conclusion}
In this paper, we first summarized the key differences between FF and NF propagation channels: the spherical wavefront assumption and the SnS assumption. Subsequently, we analyzed their impact on performance evaluation. Numerical results indicated that, based on the NF channel model, the system performance degradation caused by using FF beamforming codebook did not exceed 3 dB within most of the NF region defined by the Rayleigh distance. However, to ensure the effective coverage in the NF region, it is required to use NF beamforming codebook. Furthermore, the achievable rate loss due to the FF beam training technique did not exceed 0.3 bps/Hz with that achieved by NF beam training technique in our simulation. However, with the advancement of NF training techniques, it becomes necessary to enhance the achievable rate in NF region with only a slight increase in additional training overhead.

\section*{ACKNOWLEDGEMENT}
\label{ACKNOWLEDGEMENT}

This work was supported by the National Key R\&D Program of China (No.~2023YFB2904803), the Beijing Natural Science Foundation (L243002), the National Natural Science Foundation of China (No.~62341128), the Guangdong Major Project of Basic and Applied Basic Research (No.~2023B0303000001), the National Natural Science Foundation of China (No.~62201086) and Beijing University of Posts and Telecommunications-China Mobile Research Institute Joint Innovation Center.

\bibliographystyle{gbt7714-numerical}
\bibliography{myref}

\biographies

\begin{CCJNLbiography}{dzh.jpg}{Zihang Ding}		received the B.S. degree in communication engineering from the Beijing University of Posts and Telecommunications (BUPT) in 2023. He is currently pursuing the Ph.D. degree in electronic information engineering with BUPT. His current research interests include channel measurements and modeling, near-field communications, and array signal processing.\end{CCJNLbiography}

\begin{CCJNLbiography}{zjh.png}{Jianhua Zhang}	received the Ph.D. degree from the Beijing University of Posts and Telecommunications (BUPT) in 2003. She is currently a Professor with BUPT, the China Institute of Communications Fellow, and the Director of the BUPT–CMCC Joint Research Center. She has authored or coauthored more than 200 papers and authorized 40 patents. Her research interests include beyond 5G and 6G, AI, data mining, channel modeling for integrated sensing and communication, massive MIMO, mm-Wave, THz, visible light channel modeling, channel emulator, and OTA test. She was the recipient of several paper awards, including 2019 SCIENCE China Information Hot Paper, 2016 China Comms Best Paper, and 2008 JCN Best Paper. She was also the recipient of several prizes for her contribution to ITU–R 4G channel model (ITU–R M.2135), 3GPP relay channel model (3GPP 36.814), and 3GPP 3D channel model (3GPP 36.873). She was also a Member of 3GPP “5G channel model for bands up to 100 GHz”. From 2016 to 2017, she was the Drafting Group (DG) Chairperson of ITU–R IMT–2020 Channel Model and led the drafting of the ITU–R M. 2412 Channel Model Section. She is also the Chairwomen of the China IMT–2030 Tech Group–Channel Measurement and Modeling Subgroup and works on 6G channel model.\end{CCJNLbiography}

\begin{CCJNLbiography}{ycs.jpg}{Changsheng You}		received his Ph.D. degree from the University of Hong Kong (HKU) in 2018. He is currently an Assistant Professor at Southern University of Science and Technology. His research interests include near-field communications, intelligent reflecting surface, UAV communications, edge computing and learning. He is a Guest Editor for IEEE JSAC, an editor for IEEE TWC, IEEE COMML, IEEE TGCN and IEEE OJ-COMS. He received the IEEE Communications Society Asia-Pacific Region Outstanding Paper Award in 2019, IEEE ComSoc Best Survey Paper Award in 2021,and IEEE ComSoc Best Tutorial Paper Award in 2023. He is listed as the Highly Cited Chinese Researcher.\end{CCJNLbiography}

\begin{CCJNLbiography}{tp.jpg}{Pan Tang}		received the B.S. degree in electrical information engineering from the South China University of Technology in 2013, and the Ph.D. degree in information and communication engineering from the Beijing University of Posts and Telecommunications (BUPT) in 2019. From 2017 to 2018, he was a Visiting Scholar with the University of Southern California, Los Angeles, CA, USA. From 2019 to 2021, he was a Postdoctoral Research Associate with BUPT. He is also an Associate Professor with the State Key Laboratory of Networking and Switching Technology, BUPT. He has authored and coauthored more than 40 papers in refereed journals and conference proceedings. His research interests include millimeter wave, terahertz, and visible light channel measurements and modeling.\end{CCJNLbiography}

\begin{CCJNLbiography}{xhb.png}{Hongbo Xing}		received the B.S. degree in communication engineering from the Beijing University of Posts and Telecommunications (BUPT) in 2023. He is currently pursuing the Ph.D. degree in information and communication engineering, BUPT. His current research interests include channel modeling, XL-MIMO, near-field communication and etc.\end{CCJNLbiography}

\begin{CCJNLbiography}{yzq.jpg}{Zhiqiang Yuan}	 received his B.S. degree and Ph.D. degree from the Beijing University of Posts and Telecommunications (BUPT) in 2018 and 2024, respectively. He has been a postdoc at Southeast University since 2024. He was also a visiting Ph.D. student in the Antennas, Propagation, and Millimeter-wave Systems (APMS) section, Aalborg University, Denmark, from Jul. 2021 to Jul. 2023. His current research focuses include radio channel sounding and modeling for massive MIMO, mmWave, and THz systems, array signal processing, and channel parameter estimation.\end{CCJNLbiography}

\begin{CCJNLbiography}{mj.jpg}{Jie Meng}	received the B.S. degree in communication engineering from the Nanjing University of Posts and Telecommunications (NJUPT) in 2019. He is currently working towards the master degree in communication engineering with the Beijing University of Posts and Telecommunications (BUPT). His current research interests are wireless and mobile communication technologies. technologies.\end{CCJNLbiography}

\begin{CCJNLbiography}{lgy.png}{Guangyi Liu}	received the Ph.D. degree in circuit and system from the Beijing University of Posts and Telecommunication (BUPT) in 2006. He is the Lead Specialist and 6G Director of China Mobile. He has published more than 100 articles in referred journals and conferences and filed more than 100 patents. He was awarded “national innovation award in science and technology of 2016” of Chinese government due to his contribution to TD-LTE standardization and industrialization globally in last 10 years. In 2009 and 2013, he received two “excellent patent award” from the Chinese Intellectual property office; His current work focuses on 5G/6G research/standardization and industrialization.\end{CCJNLbiography}
\end{document}